\documentclass[aps,prl,twocolumn,superscriptaddress,showpacs]{revtex4-1}
\makeatletter

\newcommand{\Rmnum}[1]{\expandafter\@slowromancap\romannumeral #1@}
\makeatother
\usepackage{graphicx}
\usepackage{latexsym}
\usepackage{amssymb}
\usepackage{amsfonts}
\usepackage{soul}
\usepackage{color}
\usepackage{amsmath}
\usepackage{natbib}
\usepackage[english]{babel}

\begin{document}

\title{Magnetic Dipoles at Topological Defects in the Meissner state of a Nanostructured Superconductor}

\author{Jun-Yi Ge}
\affiliation{INPAC--Institute for Nanoscale Physics and Chemistry, KU Leuven, Celestijnenlaan 200D, B--3001 Leuven, Belgium}
\author {Vladimir N. Gladilin}
\affiliation{INPAC--Institute for Nanoscale Physics and Chemistry, KU Leuven, Celestijnenlaan 200D, B--3001 Leuven, Belgium}
\affiliation{TQC--Theory of Quantum and Complex Systems, Universiteit Antwerpen, Universiteitsplein 1, B--2610 Antwerpen, Belgium}
\author{Cun Xue}
\affiliation{INPAC--Institute for Nanoscale Physics and Chemistry, KU Leuven, Celestijnenlaan 200D, B--3001 Leuven, Belgium}
\affiliation{Key Laboratory of Mechanics on Disaster and Environment in Western China attached to the Ministry of Education of China, Department of Mechanics and Engineering Sciences, School of Civil Engineering and Mechanics, Lanzhou University, Lanzhou, Gansu 730000, China}
\author{Jacques Tempere}
\affiliation{TQC--Theory of Quantum and Complex Systems, Universiteit Antwerpen, Universiteitsplein 1, B--2610 Antwerpen, Belgium}
\author{Jozef T. Devreese}
\affiliation{TQC--Theory of Quantum and Complex Systems, Universiteit Antwerpen, Universiteitsplein 1, B--2610 Antwerpen, Belgium}
\author{Joris Van de Vondel}
\affiliation{INPAC--Institute for Nanoscale Physics and Chemistry, KU Leuven, Celestijnenlaan 200D, B--3001 Leuven, Belgium}
\author{Youhe Zhou}
\affiliation{Key Laboratory of Mechanics on Disaster and Environment in Western China attached to the Ministry of Education of China, Department of Mechanics and Engineering Sciences, School of Civil Engineering and Mechanics, Lanzhou University, Lanzhou, Gansu 730000, China}
\author{Victor V. Moshchalkov}\email{Victor.Moshchalkov@fys.kuleuven.be}
\affiliation{INPAC--Institute for Nanoscale Physics and Chemistry, KU Leuven, Celestijnenlaan 200D, B--3001 Leuven, Belgium}

\date{\today}

\begin{abstract}
In a magnetic field, superconductivity is manifested by total magnetic field
expulsion (Meissner effect) or by the penetration of integer multiples of
the flux quantum $\Phi_0$. Here we present experimental results revealing
magnetic dipoles formed by Meissner current flowing around artificially introduced topological
defects (lattice of antidots). By using scanning Hall probe microscopy, we
have detected ordered magnetic dipole lattice generated at spatially
periodic antidots in a Pb superconducting film. While the conventional homogeneous
Meissner state breaks down, the total magnetic flux of the magnetic dipoles
remains quantized and is equal to zero. The observed magnetic dipoles
strongly depend on the intensity and direction of the locally flowing
Meissner current, making the magnetic dipoles an effective way to
monitor the local supercurrent. We have also investigated the first
step of the vortex depinning process, where, due to the generation of
magnetic dipoles, the pinned Abrikosov vortices are deformed and shifted from their
original pinning sites.
\end{abstract}

\maketitle

Quantum and classical vortices play a crucial role in our understanding of the universe ranging from subatomic particles through superfluids \cite{Blaauwgeers,Bewley,Zwierlein,Leggett,Awschalom}, superconducting and Bose-Einstein condensates (BEC) \cite{Cornell,Nelson,Weiler,Abo-Shaeer,Dresselhaus}, exciton-polariton condensates \cite{Lagoudakis-1,Lagoudakis}, Karman vortex streets in ocean currents \cite{Ginzburg} to cosmology \cite{Zurek,Kibble}. Depending on the system and the boundary conditions under consideration, vortices as topological defects can be classical or quantum. Contrary to a classical vortex, the angular momentum in a quantum vortex carries a quantized circulation as the phase of the wave function changes around the vortex core by discrete values $2\pi L$  (with \textit{L} integer). Besides vortices, vortex-antivortex (v-av) pairs (or v-av dipoles) can appear in classical gases or fluids \cite{Couder,Engels} and also in quantum matter \cite{Moshchalkov,Roumpos,Neely,Simula,Chibotaru,Moshchalkov-2}.

The quantization of the vorticity prevents vortices to form below certain rotation speed (for neutral superfluids such as helium) or below a first critical magnetic field (for superconductors). As a result, neutral superfluids stay in the zero angular momentum state for sufficiently low rotation speeds \cite{Hess,Lynall}, 
and superconductors exhibit perfect diamagnetism at low magnetic fields \cite{Bardeen,Parks}. The latter phenomenon, discovered by Meissner and Ochsenfeld, is also known as the \textit{Meissner effect}. In type-II superconductors the Meissner state, which holds up to a critical field $\mu_0H_{\textit{c1}}$, is expected to be a uniform zero-flux state. Only when the external field is above $\mu_0H_{\textit{c1}}$, a non-uniform coexistence state with superconducting and normal regions, in the form of an Abrikosov vortex ($\Phi_0$-vortex) lattice \cite{Abrikosov}, is established. However, superconductors below the critical field are not necessarily in the zero-flux state. One example is that, in an ultra fast cooled superconducting film, spontaneous $\Phi_0$-vortices with opposite polarities can nucleate even in the absence of an external field \cite{Maniv,Golubchik}. Recently, the breakdown of the uniform Meissner state has also been discussed in the context of superconducting films where bound v-av dipoles are formed due to the flow of Meissner current \cite{Eisenmenger,Carneiro,Ge-NC}. However, due to the the random distribution of the spontaneously introduced pinning centers and the variation of their pinning strength, the interaction between pinning centers and the Meissner current and the v-av dipole itself is not well understood.
Moreover, local control and manipulation of magnetic field in superconductors is of great importance for designing information-storage superconducting electronics.
\begin{figure}[!t]
\centering
\includegraphics*[width=1\linewidth,angle=0]{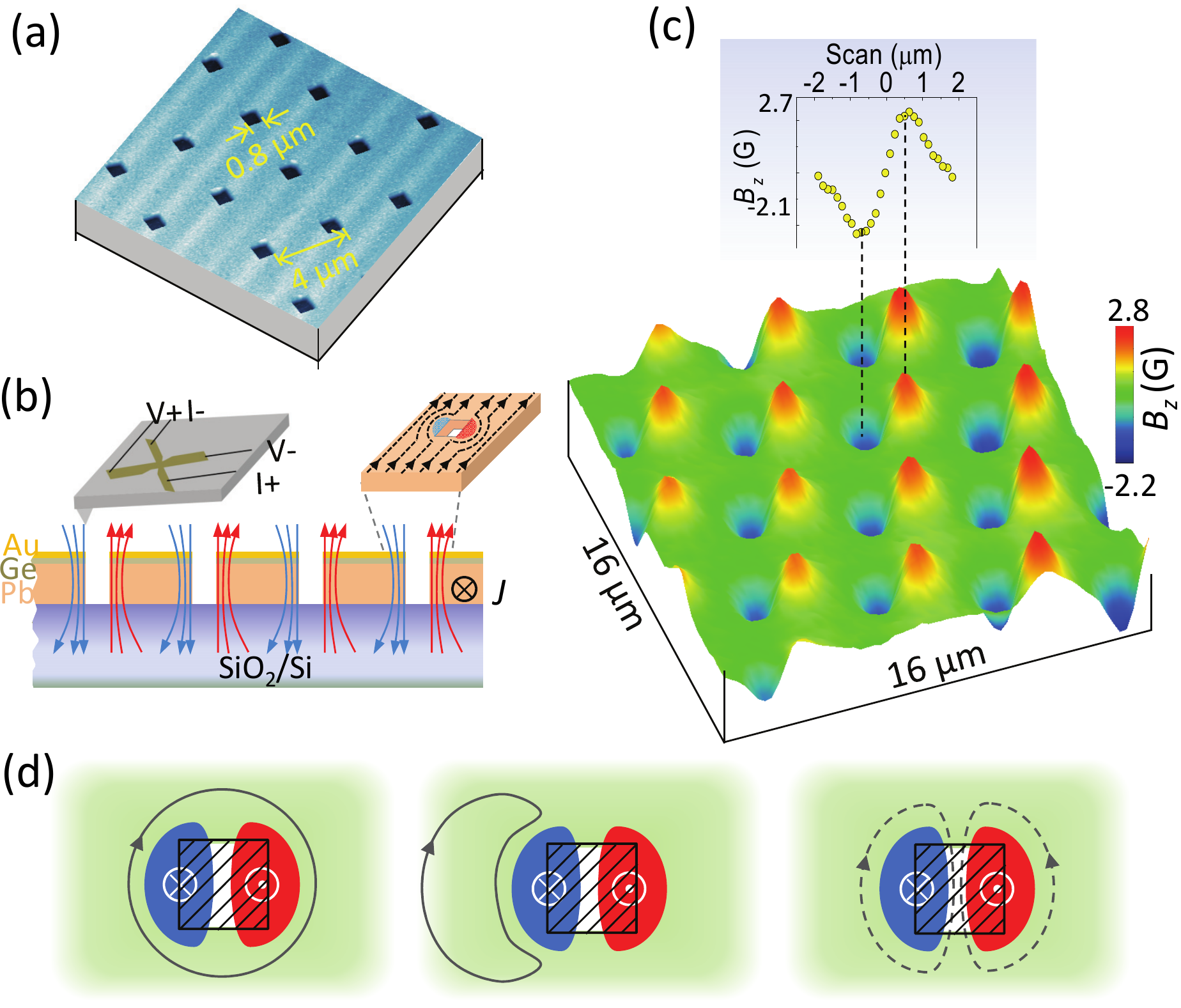}
\caption{(a) Atomic force microscopy (AFM) image of the sample structure. (b) Simplified side-view schematics of the scanning Hall probe microscopy (SHPM) measurement (not drawn to scale). The arrows indicate the local magnetic field, with opposite polarity, generated due to the redistribution of the Meissner current (top-right inset) at the antidots. (c) SHPM image of the magnetic dipole lattice measured in the Meissner state at $T=4.2$ K and $B_\textrm{0}$=6 G. One of the dipole field profiles is plotted in the top inset. (d) Different type of closed contours in a superconducting state with a magnetic dipole. The shaded area indicates the antidot which is non-superconducting. }
\label{fig1}
\end{figure}

In this Letter, we show that, in a nanostructured superconductor with a periodic array of topological defects such as prefabricated antidots, a well ordered magnetic dipole lattice can be created in the Meissner state. The magnetic flux of each pole/antipole can be well controlled by changing the supercurrent. Furthermore, we show that such a magnetic dipole lattice can be used to monitor the local current density and direction of flow. We also find that the deformation of a $\Phi_0$-vortex pinned by the antidot cell subjected to the Lorentz force can be understood by considering overlap between a magnetic dipole and a conventional $\Phi_0$-vortex. 

A topographic image of the used sample surface is shown in Fig.~1a. Square antidots are introduced with a size of $a=0.8$~$\mu$m and a period of $d=4$~$\mu$m. Antidots in superconductors are known to act as strong pinning centers at low fields \cite{Moshchalkov1998,Silva,Metlushko}. The local magnetic field distribution is mapped by low temperature scanning Hall probe microscopy (SHPM) \cite{Bending}. Our measurements revealed that up to two flux quanta can be trapped by each antidot at high enough fields. A schematic view of the SHPM experiments is shown in Fig.~1b. The Hall cross is mounted together with a scanning tunneling microscope (STM) probe tip which is used to bring the Hall sensor in close vicinity of the sample surface. An external magnetic field $B_0$, smaller than the penetration field $B_\textrm{p}$ at which $\Phi_0$-vortices enter, is applied perpendicular to the sample surface. As a result, a Meissner current flowing along the sample edges is induced to screen the magnetic field. At the position of each antidot, the Meissner current lines reorient themselves to go around the antidot, thus generating a pair of magnetic poles  with opposite polarity (inset of Fig.~1b).  As shown in Fig.~1c, a well ordered magnetic dipole lattice is directly visualized. All magnetic dipoles have the same orientation, suggesting that the Meissner current flows along the same direction in this particular scanned area. 

The magnetic dipole can be considered as two fluxoids located at a certain distance from each other. It is known that, in a superconducting condensate, the fluxoid quantization follows from the phase coherence of the macroscopic wave function along a contour encircling the vortex. However, this assumes that it is possible to find a contour that encircles the fluxoid, and that such a contour never has to go through a region of zero amplitude of the superconducting order parameter. Our current sample geometry offers a way to violate this assumption, so that the individual pole (or antipole) in the magnetic dipole no longer must be quantized (Fig.~1d). Seemingly, this is in a conceptual contradiction with the quantum nature of the superconducting condensate. However, one should keep in mind that these \textit{non-integer magnetic poles and antipoles always appear in pairs as bound dipoles}. Although the individual magnetic flux of the pole and antipole  may be non-quantized the total flux in a dipole is zero and therefore remains quantized. In other words, the opposite non-quantized classical circulations of the individual poles result in a quantized circulation for the magnetic dipole. Note that, the observed peak to valley distance ($\sim 1.2$ $\mu$m) of magnetic dipoles is slightly larger than the antidot size. This might be due to the depletion of superconductivity in the sample (at the surface defining antidots) when transported to the SHPM, thus making the actual size of antidots bigger. Our theoretical simulations (see below) have shown that the peak and valley of the magnetic dipole signal are right at the edge of the antidot. This rules out the possibility to draw a closed contour enclosing only one pole of the dipole in the superconducting region.

\begin{figure}[!t]
\centering
\includegraphics*[width=1\linewidth,angle=0]{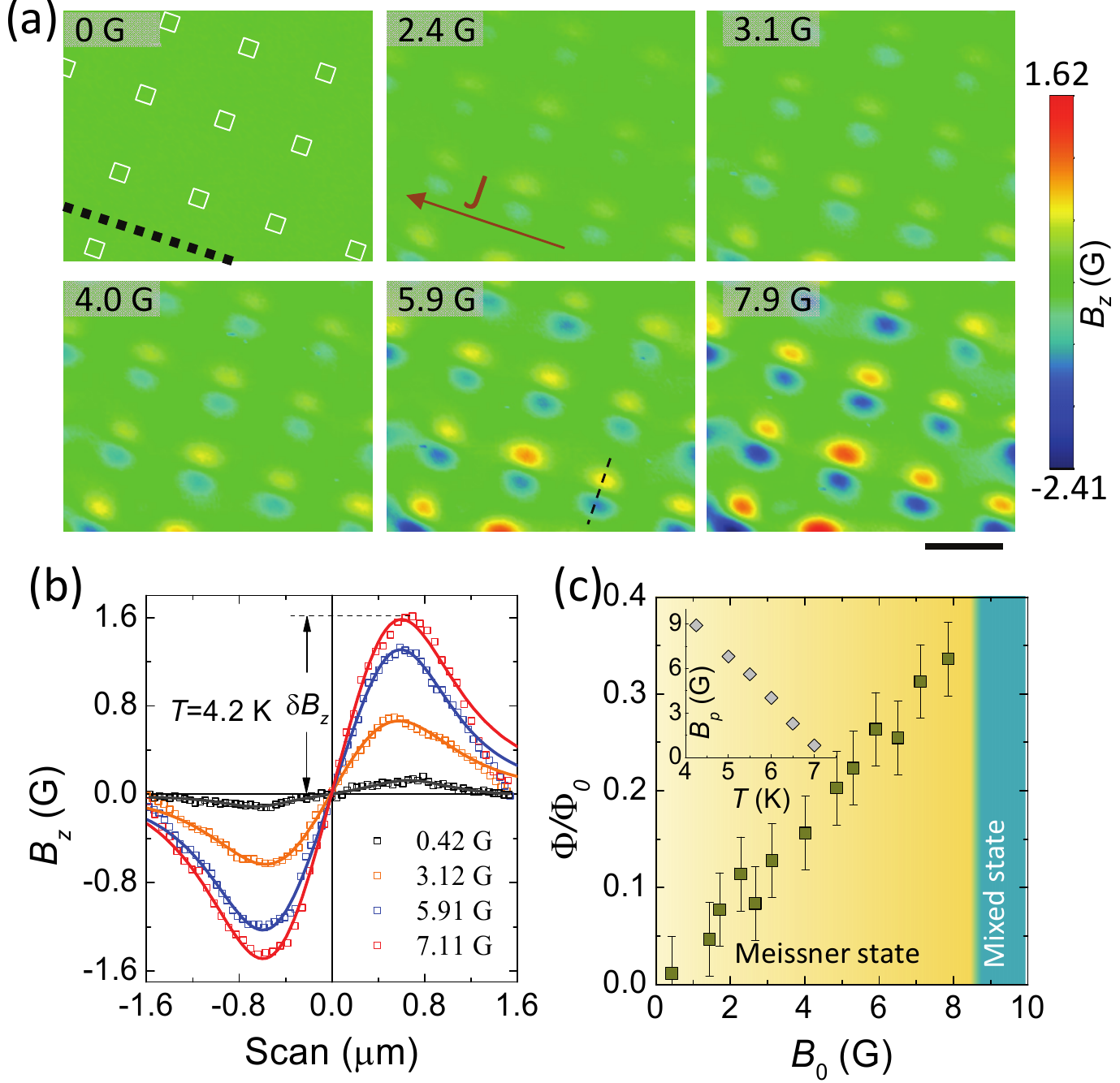}
\caption{(a) SHPM images taken after first performing zero-field cooling to 4.2 K and then increasing the external field to the value indicated in each image. The dotted line indicates the alignment of the nearest sample edge. The arrow shows the direction of flow of the Meissner current. With increasing external field, magnetic dipoles appear at the positions of the antidots marked by the squares. The scale bar equals 4~$\mu$m. (b) Magnetic field profiles of a magnetic dipole measured along the dashed line in (a), for different values of the external magnetic field. The solid lines are fitting curves with the monopole model. (c) Bound magnetic flux of each pole or antipole showing a linear dependence on the external magnetic field in the Meissner state. The inset displays the temperature dependence of the penetration field $B_\textrm{p}$, above which Abrikosov vortices enter the sample.}
\label{fig2}
\end{figure}

The magnetic flux of each pole and antipole depends on the local intensity of the flowing Meissner current. Figure~2 shows how the magnetic dipoles develop as a function of external magnetic field in the Meissner state. The scanned area is chosen close to the sample edge which is parallel to the dotted line in Fig.~2a. In the absence of external field, no magnetic dipole is observed. After applying a magnetic field,  magnetic dipoles appear at the locations of the antidots indicated by squares. A few important features of the magnetic dipoles can be mentioned. From the field profiles shown for one magnetic dipole in Fig.~2b, locally, the pole and antipole have the same absolute field intensity $\delta B_z$, which follows a linear dependence with magnetic field (see supplementary Fig.~S1). In superconductors, the magnetic field induced by fluxoid can be simulated by using the monopole model \cite{Ge,pearl2004structure,chang1992scanning,wynn2001limits}:
\begin{equation}
{B_z}(r) = \frac{\Phi }{{2\pi }}\frac{{\lambda  + {z_0}}}{{{{\left[ {{r^2} + {{\left( {\lambda  + {z_0}} \right)}^2}} \right]}^{{3 \mathord{\left/
 {\vphantom {3 2}} \right.
 \kern-\nulldelimiterspace} 2}}}}}.
\label{Monopole-1}
\end{equation}
where $B_z(r)$ is the magnetic field perpendicular to the sample surface, $r$ is the in-plane distance from the fluxoid center, $\lambda$ is the penetration depth, $z_0$ is the distance from the sample surface to the two dimensional electron gas (2DEG) of the Hall cross and $\Phi$ is the total flux carried by the fluxoid.
We have found that, as shown by the solid lines in Fig.~2b, the measured magnetic dipole field profile can be well simulated by considering two monopoles with opposite polarity which are placed at a certain distance.
In Fig.~2c we plot the field dependence of the normalized magnetic flux $\Phi/\Phi_0$ from the fitting of the data for one magnetic dipole at $T=4.2$ K. The $\Phi$ value increases linearly with external field until the first $\Phi_0$-vortex enters the sample at the penetration field $B_p$, above which the mixed state is established. \textit{This means that the magnetic flux carried by each pole or antipole in a magnetic dipole is not quantized.} This is the main difference between poles of a magnetic dipole presented here and a $\Phi_0$-vortex. As shown in our previous calculations, $\Phi_0$-vortices and antivortices can be generated when the field intensity of the magnetic dipole becomes big enough \cite{Ge-NC}. It has been reported that, close to a defect, the vortex core extends a string towards the defect edge and the circulating current will engulf the defect-vortex pair \cite{Priour}. In our case, whether the magnetic pole/antipole has a core (suppressed order parameter) is rather difficult to access, since our SHPM only measures the magnetic field distribution. From this point of view, more direct vortex core studies with scanning tunnelling microscopy might help to understand deeper this phenomenon. 

\begin{figure}[!t]
\centering
\includegraphics*[width=1\linewidth,angle=0]{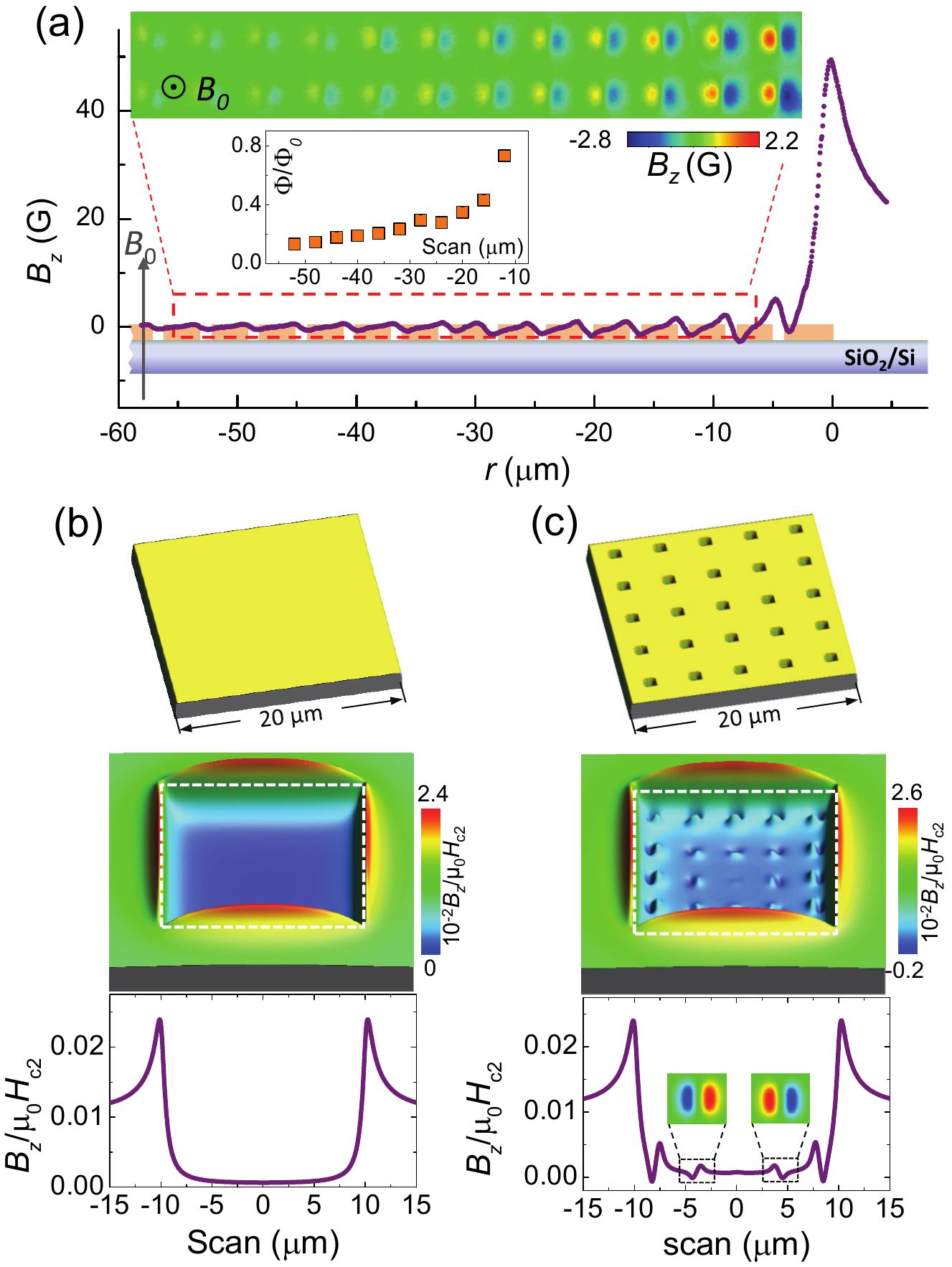}
\caption{(a) The measured magnetic field profile along the magnetic dipole distribution perpendicular to a sample edge showing the modified critical state at $T=4.2$ K and $B_0=6.3$ G. The upper inset shows the magnetic dipoles mapped in the dashed rectangle area. The lower inset shows the normalized magnetic flux of each pole (antipole) at the antidots as a function of distance $r$ from the edge. The simulated magnetic field distribution for a $20\times20$ $\mu m^2$ plain film with (c) and without (b) antidots. In both (b) and (c), the upper panels show schematically the film topography while the middle and lower panel display the calculated magnetic field distribution and the field profile crossing the centers of two opposite edges, respectively. The white dashed lines indicate the positions of film edges. }\label{fig3}
\end{figure}

Another feature is that along the direction of the Meissner current, all the magnetic dipoles exhibit the same field intensity, while in the direction perpendicular to the Meissner current, it decreases with increasing distance from the edge (Fig.~2d). This results from the distribution of the Meissner current density which decreases going away from the sample border \cite{tinkam1996introduction}.
To study in more details the phenomenon, we mapped the magnetic field distribution in the area deeper into the sample. At the edge, due to the shielding, the magnetic field is compressed, resulting in a pronounced peak at $r=0$ in the magnetic field profile shown in Fig.~3a. Inside the superconductor, magnetic dipoles can be detected up to 58~$\mu m$ from the sample edge with the field intensity decreasing continuously. The magnetic flux of each pole (antipole), determined using Eq.~(1), is plotted in the inset of Fig.~3a as a function of the distance from the sample edge clearly demonstrating non-integer flux of magnetic pole/antipole forming the magnetic dipoles at antidots. Close to the edge, the magnetic pole (antipole) carries more magnetic flux due to a relatively large local supercurrent. At 4.2 K and $B_0=6.3$ G, the maximum magnetic flux carried by a pole (antipole) reaches values up to 0.8$\Phi_0$. However, in our measurements, no $\Phi_0$-vortex is generated at interstitial positions between the pinning centers. 

Using the time dependent Ginzburg-Landau (TDGL) equations (see supplementary), we performed simulations for a superconducting film with a thickness of 100 nm and with lateral sizes 20~$\mu \textrm{m}\times20~\mu$m. The pinning centers are introduced as $0.8\times0.8~\mu$m$^2$ size antidots, arranged in a square lattice (Fig.~3c). For comparison the magnetic field distribution of a plain film is also calculated and shown in Fig.~3b. The TDGL simulations, corresponding to zero-field cooling followed by an increase of the applied magnetic field, reveal a well ordered magnetic dipole array. The orientation of the magnetic dipoles varies accordingly with changing the Meissner current direction, in agreement with the experimental observations (see supplementary Fig.~S2). Our calculations further reveal that the magnetic field intensity of the magnetic dipoles decreases rapidly with the distance $z_0$ from the sample surface (supplementary Fig.~S3). The calculated data shown in Figs.~3b and 3c correspond to $z_0=0.3$~$\mu$m, comparable to our experimental situation. Above $z_0=1$~$\mu$m, the magnetic dipoles become weak and are barely detectable.

\begin{figure}[!t]
\centering
\includegraphics*[width=1\linewidth,angle=0]{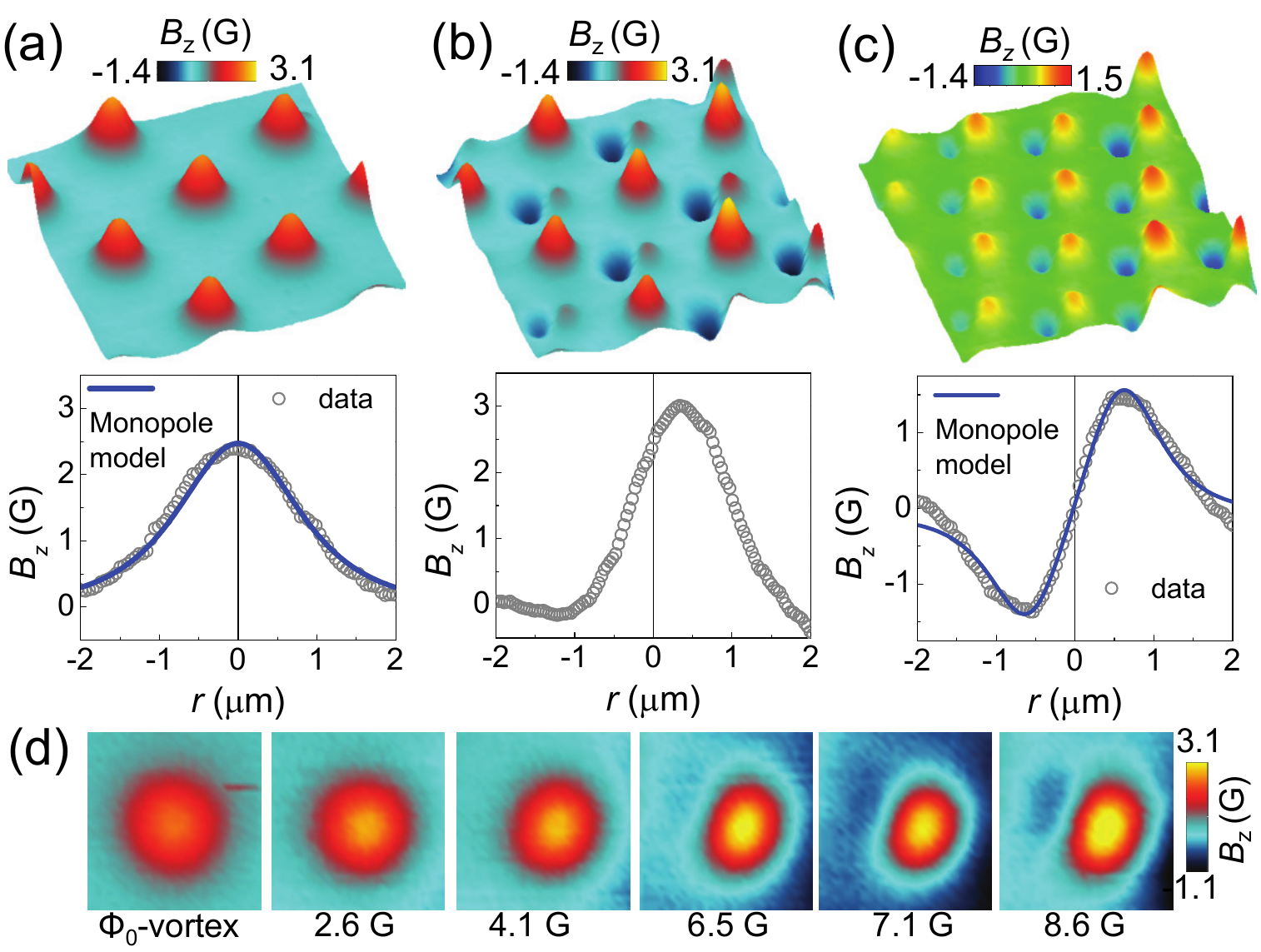}
\caption{(a) FC $\Phi_0$-vortex lattice, corresponding to half matching configuration, observed at $B_0=0.65$ G and $T=4.2$ K. (b) Coexistence of magnetic dipole lattice with $\Phi_0$-vortex lattice after increasing magnetic field from 0.65 G to 7.1 G. (c) Differential image obtained by subtracting a) from b). In (a)-(c), the typical field profiles for a $\Phi_0$-vortex, deformed $\Phi_0$-vortex and the magnetic dipoles are shown below each image. The solid line is the fitting curve with the monopole model. (d) SHPM images showing the progressive deformation of a $\Phi_0$-vortex with increasing the external magnetic field as indicated.}\label{fig4}
\end{figure}

In superconductors, vortices experience a Lorentz force $\vec{F_\textrm{L}}=\vec{J}\times \vec{B}$ exerted by the current. Once the Lorentz force overcomes the attractive force between pinning centers and vortices, vortices are unpinned and the superconductor becomes resistive due to the vortex motion. However, the depinning process itself has rarely been studied experimentally, especially within a single vortex resolution. Using SHPM, we have analyzed the effect of supercurrent on the pinned $\Phi_0$-vortex lattice. Figure~4a shows the $\Phi_0$-vortex lattice observed after cooling at half matching field. Exactly half of the antidots are occupied by the $\Phi_0$-vortices with the field profile being well simulated by the monopole model.

With increasing external field, a shielding current is induced that prevents penetration of new vortices into the sample. As a result, magnetic dipoles which can clearly be seen at the unoccupied antidots in Fig.~4b are generated, leading to the coexistence of a $\Phi_0$-vortex lattice and a magnetic dipole lattice. Furthermore, we observe that the pinned $\Phi_0$-vortices are moved away from their original positions at the antidots and that their intensity is dramatically enhanced, especially close to the sample edge. By subtracting the field distribution of Fig.~4a from that in Fig.~4b, a well ordered magnetic dipole lattice is observed as shown in Fig.~4c. Note that the spacing of the dipoles in Fig.~4c is exactly two times smaller that the distance between the dipoles visible in Fig.~4b. Invisible dipoles in Fig.~4b are merged with the fields of $\Phi_0$-vortex lattice. This suggests that the deformation of pinned $\Phi_0$-vortices in a superconductor mainly arises from the locally generated magnetic dipoles at the pinning centers. 
The evolution of the $\Phi_0$-vortex deformation is shown in detail in Fig.~4d. Clearly, as expected, the $\Phi_0$-vortex becomes elongated along the flow direction of the current as the external field increases. Our results provide direct visualization of the first step of the vortex depinning process. This provides a new perspective way to design effective pinning centers, where along the direction of flowing current, magnetic dipoles should be minimized (see supplementary Fig. s4).

We have presented direct experimental evidence of magnetic dipoles generated at artificial pinning centers in the Meissner state of a superconducting film. We show that these magnetic dipoles form well ordered lattices, in which the bound magnetic flux of each pole (antipole) can be tuned by changing the flowing current. Each magnetic pole or antipole is not quantized. Note, however, that the total magnetic flux of each dipole consisting of bound pole-antipole remains integer (zero), which is in full agreement with the quantum nature of superconductivity. 
The possibility to obtain magnetic dipoles, where the constituent magnetic poles themselves are not quantized, is also important for the study of quantum turbulence \cite{Henn}, in particular, in ultracold quantum gases, which have great potential as quantum simulators \cite{Romero-Isart,Weiss}. Moreover, such magnetic dipoles have been shown to cause $\Phi_0$-vortex deformation. Further studies, for instance, an analysis of the order parameter distribution in the presence of magnetic dipoles by using scanning tunneling microscopy, would help to reveal new facets of this phenomenon.

We acknowledge the support from the Methusalem funding by the Flemish government, the Flemish Science Foundation (FWO-Vl) and the MP1201 COST action. J.T. also acknowledges support from the Research Council of Antwerp University (BOF). Y.Z. and C.X. acknowledge the National Natural Science Foundation of China (No. 11421062) and the National Key Project of Magneto-Constrained Fusion Energy Development Program (No. 2013GB110002). C.X. also acknowledge the CSC program.


\end{document}